\documentclass[preprint]{aastex}

\shorttitle{Black Holes and Galactic Bulges} 
\shortauthors{F. C. Adams}

\newcommand{\be}{\begin{equation}}
\newcommand{\ee}{\end{equation}}
 
\newcommand{\mbh}{{ M_{\rm bh}}} 
\newcommand{\jbh}{{ J_{\rm bh}}} 

\newcommand{\tstar}{{ \tau_\ast }} 
\newcommand{\tblg}{{ \tau_{\rm blg} }} 
\newcommand{\tbh}{{ \tau_{\rm bh} }} 
\newcommand{\tcross}{{t_{\rm cross}}}
\newcommand{\fa}{{ {\cal F}_A }} 
\newcommand{\fdm}{{ {\cal F}_{DM} }} 
\newcommand{\thb}{{ \theta_{\rm bh} }} 
\newcommand{\thbulg}{{ \theta_B }} 
\newcommand{\acon}{{ {\widetilde A} }} 
\def\kms{\ifmmode \hbox{ \rm km s}^{-1} \else{ km s$^{-1} $}\fi}

\def\half{{1 \over 2}}
\def\myr{\ifmmode \hbox{\rm Myr}\else{Myr}\fi} 
\def\gyr{\ifmmode \hbox{\rm Gyr}\else{Gyr}\fi} 
\def\kpc{\ifmmode \hbox{\rm  kpc}\else{kpc}\fi} 
\def\Msun{M_\odot}

\newcommand{\rsw}{ R_S } 
\newcommand{\mrat}{{ \mu_B }} 
\newcommand{\aeff}{ a_{\rm eff} } 

\begin{document}

\title{Formation of Supermassive Black Holes in Galactic Bulges: \\ 
A Rotating Collapse Model Consistent with the $\mbh-\sigma$ Relation} 

\author{Fred C. Adams$^{1,2}$, David S. Graff$^{2,3}$, Manasse Mbonye$^1$, 
and Douglas O. Richstone$^{1,2}$} 
 
\affil{$^1$Michigan Center for Theoretical Physics \\
Physics Department, University of Michigan, Ann Arbor, MI 48109}

\affil{$^2$Astronomy Department, University of Michigan, Ann Arbor, MI 48109}

\affil{$^3$Department of Math and Science \\ 
United States Merchant Marine Academy, Kings Point, NY 11024 }  

\affil{DRAFT: 6 March 2003} 

\begin{abstract} 

Motivated by the observed correlation between black hole masses $\mbh$
and the velocity dispersion $\sigma$ of host galaxies, we develop a
theoretical model of black hole formation in galactic bulges (this
paper generalizes an earlier ApJ Letter). The model assumes an initial
state specified by a a uniform rotation rate $\Omega$ and a density
distribution of the form $\rho = \aeff^2 / 2 \pi G r^2$ (so that
$\aeff$ is an effective transport speed).  The black hole mass is
determined when the centrifugal radius of the collapse flow exceeds
the capture radius of the central black hole (for Schwarzschild
geometry). This model reproduces the observed correlation between the
estimated black hole masses and the velocity dispersions of galactic
bulges, i.e., $\mbh \approx 10^8 M_\odot (\sigma/200 \, {\rm km \,
s^{-1}})^4$, where $\sigma = \sqrt{2} \aeff$. To obtain this
normalization, the rotation rate $\Omega \approx 2 \times 10^{15}$
rad/s. The model also defines a bulge mass scale $M_B$. If we identify
the scale $M_B$ with the bulge mass, the model determines the ratio
$\mrat$ of black hole mass to the host mass: $\mrat$ $\approx$ 0.0024
$(\sigma/200 \, {\rm km \, s^{-1}})$, again in reasonable agreement
with observed values. In this scenario, supermassive black holes form
quickly (in $\sim10^5$ yr) and are born rapidly rotating (with $a/M
\sim 0.9$).  This paper also shows how these results depend on the
assumed initial conditions; the most important quantity is the initial
distribution of specific angular momentum in the pre-collapse state.

\end{abstract}


\section{INTRODUCTION} 

During the past decade, the observational evidence for massive black
holes has crossed a threshold of reliability and black holes are now
considered to be discovered. Almost every galaxy is thought to harbor
a supermassive black hole anchoring its center (e.g., Magorrian et al.
1998; Kormendy \& Richstone 1995). Our own Milky Way galaxy contains a
modest central black hole, with its mass estimated at $M_{\rm mw}$
$\approx$ $3 \times 10^{6}$ $M_\odot$ (e.g., Genzel et al. 1996, Ghez
et al. 1998). The properties of these black holes and their
connections to their galactic hosts are currently the subject of
intensive investigation.

A striking aspect of the black hole-galaxy connection has been
recently reported. Two competing groups have observed a relationship
between the velocity dispersion $\sigma$ of the host galaxy and the
mass $\mbh$ of its central (supermassive) black hole (Gebhardt et
al. 2000; Ferrarese \& Merritt 2000). This correlation can be written
in the form 
\be 
\mbh = M_0 \, (\sigma/200 \kms)^\gamma \, , 
\label{eq:observed} 
\ee
where the two observational teams find slightly different values for
the constants in this scaling relation. The exact values derived from
the data depend on the fitting procedure and are sensitive to the
exclusion of outlying points (see Merritt \& Ferrarese 2000, 2001). A
recent in-depth analysis (Tremaine et al. 2002) finds that $\gamma$ =
4.02 $\pm$ 0.32 with a mass scale $M_0$ = $1.3 \times 10^8 M_\odot$. 
In any case, the observed correlation is remarkably tight: The
observed scatter in black hole mass $\mbh$ at fixed dispersion
$\sigma$ is less than 0.30 dex (about a factor of 2).  Furthermore,
the observed relation appears to be independent of the Hubble type,
profile type, or galactic environment. Previous observational surveys
have found correlations between the black hole mass and bulge
luminosity (Richstone et al. 1998; Magorrian et al. 1998; van der
Marel 2000; Kormendy \& Richstone 1995; see also Carollo, Stiavelli,
\& Mark 1998), but the relation [\ref{eq:observed}] appears to be far
more robust. This observed scaling relationship provides an important
constraint on theories of galaxy formation and bulge formation. Such
theories must ultimately account for the production of supermassive
black holes at galactic centers and the observed relationship between
black hole mass and galactic velocity dispersion.

In a previous paper (Adams, Graff, \& Richstone 2001; hereafter Paper
I), we presented a theoretical model for black hole formation during
the collapse and formation of galactic bulges. This model uses an
idealized treatment to describe the collapse of the inner part of
protogalaxies. The initial state is assumed to have a density
distribution of the form $\rho \propto r^{-2}$ (like that of a
singular isothermal sphere) and a uniform rotation rate; the initial
conditions are characterized by an effective transport speed $\aeff$
and the rotation rate $\Omega$. As the collapse develops, material
falls inward from ever larger starting radii and carries larger
amounts of specific angular momentum.  The black hole mass is
determined when the centrifugal radius of this collapse flow exceeds
the capture radius of the black hole growing at the center.  In spite
of its idealized nature, this simple model correctly accounts for the
observed $\mbh - \sigma$ relation (equation [\ref{eq:observed}]) with
no free adjustable parameters and also predicts the observed ratios of
black hole mass to the host (bulge) mass scale. Because of this
preliminary success, the model deserves further exploration, which is
the subject of this present work.

We note that many other theoretical models have been developed to
explain the observed relationship between hole mass and galactic
velocity dispersion (e.g., see the recent review of Richstone 2002).
A semi-analytic model of merger-driven starbursts with black hole
accretion (Haehnelt \& Kauffman 2000; Kauffman \& Haehnelt 2000)
provides a correlation of the observed form (with the proper choice of
the free model parameters). Several models are based on the idea that
black hole accretion can influence star formation and gas dynamics in
the host galaxy; this feedback can occur through ionization,
mechanical work, and heating (e.g., Ciotti \& Ostriker 1997, 2001;
Blandford 1999; Silk \& Rees 1998).  The model of Blandford (1999)
predicts that $\mbh < \eta \, \sigma^{5}$ whereas the model of Silk \&
Rees (1998) implies $\mbh \propto \sigma^5$. The idea that the black
hole mass is limited by disk accretion has been explored by Burkert \&
Silk (2001). Before the observational correlation was discovered,
Daniel \& Loeb (1995) argued that the seeds for quasar black holes
could originate from the collapse of low angular momentum regions.
Finally, the accretion of collisional dark matter indicates a scaling
relation of the form $\mbh \propto \sigma^{4 - 4.5}$ (Ostriker 2000).

This paper is organized as follows. In \S 2, we review and extend the
model presented in Paper I, and describe the orbital infall solutions
in greater detail; we also generalize the model to include continued
infall onto the black hole at late times.  In \S 3, we consider mass
accumulation onto the black hole through disk accretion, more general
initial conditions, the effects of mergers, and nonzero quadrapole
moments in the initial conditions.  We conclude, in \S 4, with a
summary and discussion of our results.

\section{THE ROTATING COLLAPSE MODEL} 

In this section, we review and expand upon the basic model of black
hole formation during the collapse of a forming galaxy (see Paper I).
In this context, we examine the collapse of the inner part of a region
that will ultimately form the bulge of a galaxy.

\subsection{Initial Conditions} 

The calculation starts at the time of maximum expansion for the main
body of the bulge. The main characteristics of the model can be
summarized as follows:

[1] All of the matter participating in the collapse -- including
baryons, dark matter, and stars -- are unsegregated. In particular, we
assume that all of the collapsing matter has the same initial
distribution of specific angular momentum, as this distribution is the
key ingredient in producing black holes with the observed properties.
However, this model does allow for the possibility that additional
material, perhaps some of the dark matter, does not participate in the
collapse (see below for further discussion).

[2] The mass and density distributions in this region take the form of
a singular isothermal sphere even though the system is not in virial
equilibrium. More specifically, the initial density and mass 
distributions are assumed to have the form 
\be 
\rho(r) = { \aeff^2 \over 2 \pi G r^2 } \qquad {\rm and} \qquad 
M (r) = {2 \aeff^2 \over G} \, r  \, . 
\label{eq:initial} 
\ee
The effective transport speed $\aeff$ that specifies the initial
conditions is related to the isotropic velocity dispersion $\sigma$
according to $\sigma = \sqrt{2} \aeff$, where $\sigma$ is the velocity
dispersion of the final state. This relation results from converting
half of the original potential energy into kinetic energy, with the
overall radius of the structure shrinking by a factor of 2 (see also
below). For gaseous material, the transport speed $\aeff$ plays the
role of the sound speed. For any dark matter participating in the
collapse, the intrinsic velocity distribution of its initial state is
highly ordered with width $\delta v$ $\ll$ $\aeff \sim \sigma$ (by 
assumption; see also below). 

[3] This region is slowly rotating like a solid body (e.g., due to
tidal torques) at a well-defined initial angular frequency $\Omega$. 
In this starting configuration, both dark matter particles and parcels
of baryons that are initially located at radius $r_\infty$ have
initial angular momentum $j$ = $\Omega (r_\infty \sin \theta_0)^2$,
where $\theta_0$ is the (initial) polar angle in spherical coordinates.

[4] The central region of the collapse flow successfully produces a
``seed'' black hole in the earliest phases of evolution. The mass of
this starting black hole may be much smaller than the large ($\mbh
\sim 10^8 M_\odot$) black holes of the final states.  Once a black
hole has condensed out of the galactic center, it will grow according
to the calculations of this model. The initial seed black hole could
form by the collapse of the densest (central) part of the perturbation
or could be primordial.

The initial state is characterized by two physical variables: $\aeff$
and $\Omega$. We stress that these quantities are not free adjustable
parameters, but rather can be specified -- or at least constrained --
by observations.  First, we note that the initial transport speed
$\aeff$ is directly related to (but not equal to) the final velocity
dispersion of the final system. In collapsing regions with no
dissipation, the virial theorem implies that the scale length of the
mass distribution drops by a factor of 2 from the point of maximum
expansion (see Binney \& Tremaine 1987). Observational considerations
(e.g., the ``flat rotation curve conspiracy'') suggest that
dissipation does not greatly alter the final value of the dispersion
$\sigma$. This argument implies that the observed velocity dispersion
$\sigma$ is related to the initial transport speed $\aeff$ of the
protogalactic material through the relation $\sigma^2 = 2 \aeff^2$.

Throughout this paper, we adopt a fiducial value for the starting
rotation rate $\Omega = 6 \times 10^{-2} \myr^{-1}$ = $2 \times
10^{-15}$ rad s$^{-1}$, which ultimately provides the observed
normalization for the $\mbh - \sigma$ relation.  The scatter about
this fiducial value will produce a corresponding scatter in the
theoretical $\mbh - \sigma$ correlation. Although the initial rotation
rates of the inner portions of galaxies in their pre-collapse states
are no longer observable, this adopted value is in reasonable
agreement with expectations.  For a ballpark estimate, we can consider
the fundamental plane (Binney \& Merrifield 1998), which provides a
relationship between the half-light radii of galactic bulges and the
corresponding velocity dispersions.  For a typical value of the
velocity dispersion $\sigma = 200\kms$, the effective radius $R_E$ of
a galaxy on the fundamental plane is about 3.5 \kpc. In order of
magnitude, we expect that $\Omega \sim \sigma/R_E \sim 2 \times
10^{-15}$ rad/s.  Rotation rates of this order of magnitude are also 
consistent with those expected from observed rotational velocities 
in galactic bulges (see, e.g., Figure 4.6 of Binney \& Treamine 1987; 
Binney \& Merrifield 1998; Jarvis \& Freeman 1985; Wyse \& Gilmore 1992). 

It is useful to compare our assumptions to the typical value of the
spin parameter $\lambda$ for protogalaxies predicted by numerical
simulations (where the angular momentum originates from cosmological
torques -- see Peebles 1993). Following Bullock et al. (2001), we
write the spin parameter in the (slightly non-standard) form 
\be
\lambda = {J/M \over \sqrt{2} R V_{cir} } \, , 
\label{eq:spindef} 
\ee
where $J/M$ is the specific angular momentum, $R$ is the outer radius,
and $V_{cir}$ is the circular speed of the protobulge structure. For
our assumed initial density distribution [\ref{eq:initial}] with solid
body rotation, the specific angular momentum $j = J/M = (2/9) \Omega
R^2$ and the circular speed at $R$ is given by $V_{cir}$ = $\sqrt{2}
\aeff$. Using these results in the definition [\ref{eq:spindef}], we
find that $\lambda$ = 1/9 for our assumed initial condition. For
comparison, the values of $\lambda$ that are predicted for dark matter
halos by numerical studies of structure formation have a mean value
$\lambda = 0.04 - 0.05$ (Peebles 1993; Barnes \& Efstathiou 1987;
Bullock et al. 2001). Although these values are a factor of 2 smaller
than used here, they are evaluated for halo mass scales that are
thousands of times larger than the protobulge mass scales used here. 
Notice also that if the bulge forms via collapse with dissipation, 
then the bulge spin parameter can be larger than the halo spin 
parameter (e.g., see White 1996). 

We can think of the initial conditions $(\aeff, \Omega)$ in two
different ways: First, we can consider the effective transport speed
$\aeff$ and the rotation rate $\Omega$ as adjustable parameters that
can be varied in order to explain four quantities: the velocity
dispersion $\sigma$, the bulge size scale $R$, the bulge mass scale
$M_B$, and the central black hole mass $\mbh$. On the other hand, we
can use $\sigma$ and $R$ to specify the initial parameters $\aeff$ and
$\Omega$. In this latter case, we are left with a theory containing no
adjustable parameters, but the theory must still correctly account for
the bulge mass scale $M_B$ and the black hole mass $\mbh$ as a
function of $\sigma$.

As we show below, all material with initial conditions given by
equation [\ref{eq:initial}] follows a ballistic trajectory as it falls
toward the central black hole. This result holds for gas, stars, and
dark matter. Gas naturally takes on a centrally concentrated
distribution and approaches the form of equation [\ref{eq:initial}];
the gas density obtains this form for the limiting case of hydrostatic
equilibrium with an isothermal equation of state. However, the dark
matter is somewhat less likely to assume this same starting state.
Unlike gas particles, individual dark matter particles can have high
angular momentum even if they are part of a larger structure with low
(or zero) angular momentum. In other words, it is possible for the
dark matter to display the overall density distribution of equation
[\ref{eq:initial}], and for the structure as a whole to rotate slowly,
and still have the individual particles possess too much angular
momentum to be captured.  In order for dark matter to fall into the
central black hole, the particles must be extremely cold (with
internal velocity dispersion $\delta v$ $\ll$ $\aeff \sim \sigma$),
and the large scale streaming velocities $V_S$ in nonradial directions
must also be small ($V_S$ $\ll$ $\aeff \sim \sigma$). 
 
Observations have shown that the central black holes of galaxies are
strongly correlated with the properties of galactic bulges, but are
more weakly correlated with other galactic components such as the disk
or the dark matter halo. For example, the galaxy M33 has no bulge (but
has both a disk and a dark matter halo), but does not contain a black
hole.  Although galactic bulges may contain some dark matter, they are
probably not dominated by dark matter (e.g., Gerhard et al. 2001).
The micro-lensing optical depth of our own Milky Way bulge is high,
$\sim 2 \times 10^{-6}$, indicating that it contains many baryons in
the form of stars, stellar remnants, and brown dwarfs (e.g., Evans \&
Belokurov 2002). In fact, the baryonic content of our bulge is so high
that it is difficult to construct dynamical bulge models using all of
the baryons; very little of the mass budget is left over for a
significant dark matter component.  Dark matter may not play a
dominant role in our galactic bulge, and, by extension, may not 
dominate the determination of the observed $\mbh - \sigma$ relation.
 
\subsection{Classical Orbit Solutions} 

As the collapse proceeds, particles in the initial distribution fall
towards the galactic center.  Because the dynamical time scales
monotonically increase with radius, infalling shells of material do
not cross. The mass contained inside a given spherical shell, which
marks a particle's location, does not change as the particle falls
inward and hence orbital energy is conserved. In the classical
(nonrelativistic) regime, the orbital energy is given by 
\be
E = \half v_r^2 + \half {j^2 \over r^2} - {GM \over r} \, . 
\label{eq:energy} 
\ee 
In this problem, we consider orbits that begin their collapse
trajectories at large radii and then fall a long way toward the center
of the galaxy.  As a result, we can idealize these trajectories as
zero energy orbits.

For a given gravitational potential, we find the orbital solutions for
material falling towards the galactic center; the same orbits apply to
stars, dark matter, and parcels of gas.  In our initial calculation
(Paper I), the inner solution is derived using the gravitational
potential of a point source. This form is only used in the innermost
regime of the collapse flow where the potential is dominated by the
forming black hole. In other words, this orbital solution derived here 
is valid over the range of length scales 
\be
\rsw \ll r \ll r_\infty \, , 
\ee
where $\rsw$ is the Schwarzschild radius of the black hole and is 
given by 
\be
\rsw = {2 GM \over c^2 } \, . 
\label{eq:rschwarz} 
\ee
In general, the black hole produced through this collapse process will
be rotating so that its event horizon is not completely specified by
the Schwarzschild radius $\rsw$. Notice that at late times, long after
black hole formation is complete, dark matter and stars will miss the
black hole and continue to trace through their orbits back out toward
large radii; this behavior leads to an extended mass distribution and
the potential is no longer well described by a point potential.  At
these later times of evolution, our solution loses its validity. In
the present context, however, we only use the solution during the
early phases in which the gravitational potential of the central
region is dominated by the black hole. In addition, we note that
relativistic corrections become important as $r\to\rsw$. 

Since this potential is spherically symmetric, angular momentum is
conserved and the motion is confined to a plane described by the
coordinates $(r, \phi)$; the radius $r$ is the same in both the plane
and the original spherical coordinates. The angular coordinate $\phi$
in the plane is related to the angle in spherical coordinates by the
relation $\cos \phi$ = $\cos \theta$ / $\cos \theta_0$, where
$\theta_0$ is the angle of the asymptotically radial streamline (see
below).  For zero energy orbits, the equations of motion imply a cubic
orbit solution, 
\be
1 - {\mu \over \mu_0} = (1 - \mu_0^2) \, 
{j_\infty^2 \over G M r} \, \equiv \zeta (1 - \mu_0^2) \, , 
\label{eq:orbit} 
\ee
where $j_\infty$ is the specific angular momentum of particles
currently arriving at the galactic center along the equatorial plane.
The second equality serves to define the parameter $\zeta$.  Here, the
trajectory that is currently passing through the position ($r$, $\mu
\equiv \cos \theta$) initially made the angle $\theta_0$ with respect
to the rotation axis (where $\mu_0$ = $\cos \theta_0$).

For a given angular momentum, we can use equation [\ref{eq:energy}] to
determine the pericenter $p$, which marks the distance of closest
approach for a parabolic (zero energy) orbit.  Our assumption of
uniform initial rotation implies that $j_\infty = \Omega r_\infty^2
\sin^2 \theta_0$, where $r_\infty$ is the starting radius of the
material that is arriving at the center at a given time. The
pericenter can be written in the form 
\be
p = {j^2 \over 2GM} = 
{(r_\infty \sin\theta_0)^4 \Omega^2 \over 2GM} 
= {(GM)^3 \Omega^2 \sin^4 \theta_0 \over 2^5 \aeff^8} \, . 
\label{eq:peri} 
\ee 
In the final equality, we have used $M = M(r)$ as a label for
$r_\infty$ by inverting the mass distribution of the initial state
(equation [\ref{eq:initial}]) to find $r_\infty$=$GM/2\aeff^2$.  As in
star formation theory (Shu et al. 1987; Cassen and Moosman 1981), we
define the centrifugal radius $R_C$ of the flow according to 
\be
R_C = {\Omega^2 G^3 M^3 \over 16\aeff^8} \, , 
\label{eq:rcent} 
\ee
which represents the radius of a circular orbit with angular momentum
$j_\infty$ for incoming matter falling within the equatorial plane.
For motion in the equatorial plane, this radius $R_C$ is twice as
large as the pericenter $p$ for a parabolic orbit with the same
angular momentum.

Given the orbital solution (equation [\ref{eq:orbit}]), 
we can find the velocity fields for the collapse flow, 
\be
v_r = - \Biggl( {G M \over r} \Biggr)^{1/2}
\Bigl\{ 2 - \zeta (1 - \mu_0^2) \Bigr\}^{1/2} \, , 
\label{eq:vrad} 
\ee
\be
v_\theta = \Biggl( {G M \over r} \Biggr)^{1/2}
\Biggl\{ {1 - \mu_0^2 \over 1 - \mu^2 } \, (\mu_0^2 - \mu^2) \, 
\zeta \Biggr\}^{1/2} , 
\label{eq:vtheta} 
\ee
\be
v_\varphi = \Biggl( {G M \over r} \Biggr)^{1/2} \, (1 - \mu_0^2) 
\, ( 1 - \mu^2 )^{-1/2} \, \zeta^{1/2} \, . 
\label{eq:vazimuth} 
\ee 
Since $\zeta$, $\mu$, and $\mu_0$ are related through the orbit
equation [\ref{eq:orbit}], the velocity field is completely determined for 
any position $(r, \theta)$.  

The density distribution of the infalling material can be obtained
by applying conservation of mass along a streamtube (Terebey, Shu,
\& Cassen 1984; Chevalier 1983), i.e.,
\be 
\rho(r,\theta) \, v_r \, r^2
\sin\theta \, d\theta \, d\varphi = - { {\dot M} \over 4 \pi}
\sin\theta_0 \, d\theta_0 \, d\varphi_0 , 
\ee 
where $\dot M$ is the total rate of mass flow (inward) through a
spherical surface (e.g., Shu 1977).  Combining the above equations, 
we can write the density profile of the incoming material in the form 
\be 
\rho(r,\theta) = { {\dot M} \over 4 \pi |v_r| r^2 }
{d \mu_0 \over d \mu} \, . 
\label{eq:density} 
\ee
The properties of the collapsing structure determine the orbit
equation [\ref{eq:orbit}], which in turn determines the form of
$d\mu_0/d\mu$. With the radial velocity given by equation
[\ref{eq:vrad}], the density field is thus completely specified
in analytical form (implicitly). 

Notice that we have ignored gas pressure in the collapse solution.
Dark matter and stars always exhibit pressure-free (ballistic)
behavior and our approximations are automatically justified for these
components. Even for gas, however, the collapse flow always approaches
a pressure-free form in the inner region. This (somewhat remarkable) 
characteristic follows from considering the gaseous portion of the
collapse flow to be a scaled-up version of the collapse flows that
have been studied previously for star formation theories (Shu 1977;
Terebey, Shu, \& Cassen 1984; see also Adams 2000). In this case, the
collapse of the initial state (with density distribution
[\ref{eq:initial}]) proceeds from inside-out and gas parcels in the
central portion of the flow always approach ballistic trajectories.

\subsection{The Mass Scale for Galactic Black Holes} 

With the collapse solution in place, we can estimate the mass of
forming black holes. The collapse flow defines a critical mass scale
$M_C$, which will be roughly comparable to the final black hole mass.
In the earliest stages of collapse, incoming material falls to small
radii $r < \rsw$, where $\rsw$ is the Schwarzschild radius of the
forming black hole. The mass that determines $\rsw$ is the total mass
$M={\dot M}t$ that has fallen thus far, i.e., we assume that the black
hole mass $\mbh$ = $M$ in this early evolutionary stage.  As the
collapse proceeds, incoming material originates from ever larger radii
and carries a commensurate increase in specific angular momentum. The
centrifugal barrier of the collapse flow thus grows with time.

If the pericenter $p$ is sufficiently small, ballistic particles will
pass inside the horizon of the black hole and be captured; even
particles that pass close to the horizon will be captured (Misner,
Thorne, \& Wheeler 1973; hereafter MTW).  As mass accumulates in the
black hole, its horizon scale and capture radius grow linearly with
mass.  The pericenter of particles in ballistic orbits, falling from
our assumed mass distribution, increases as $p \propto r_\infty^3 \sim
M^3$ (equation [\ref{eq:peri}]).  In the earliest stages of the
collapse, all of the falling material is captured by the black hole.
Later, this growth mechanism tapers off when the black hole mass
reaches a critical point defined by equating the pericenter $p$ (for
$\theta = \pi/2$ orbits) to the capture radius of the black hole. In
Schwarzschild geometry, particles coming inwards from infinity on zero
energy orbits are captured by the black hole if $p < 4 \rsw$ (MTW),
where $\rsw$ is the Schwarzschild radius (equation [\ref{eq:rschwarz}]).
The condition $p=4 \rsw$ defines the critical mass scale $M_C$, 
\be
M_C \equiv { 16 \aeff^4 \over G c \Omega} \, =
{ 4 \sigma^4 \over G c \Omega} \, ,  
\label {eq:mcrit} 
\ee 
which represents the mass at which direct accretion onto the black
hole becomes compromised. In the original version of our model (Paper
I), this critical mass scale $M_C$ determined the observed black hole
mass $\mbh$.  Notice that equation [\ref{eq:mcrit}] displays the
correct (observed) scaling with velocity dispersion $\sigma$.  

Even after the centrifugal barrier grows larger than the Schwarzschild
radius, however, the black hole can gain additional mass from material
falling on streamlines that are oriented along the rotational poles of
the system. After the critical point (described above) is reached, the
fraction of the infalling material that lands at such small radii is a
relatively rapidly decreasing function of time. As a result, this
effect makes the black hole mass larger by a modest factor $\fa$,
where the maximum value $\fa$ $\approx$ 1.35, as we calculate next:

The mass infall rate ${\dot M}_{\rm bh}$ for material falling directly 
onto the black hole itself is given by 
\be
{\dot M}_{\rm bh} = \int_0^1 d\mu \, 4\pi (\alpha \rsw)^2 |v_r| 
\rho(\alpha \rsw, \mu) \, , 
\ee
where $\mu = \cos \theta$. We evaluate the density at the capture
surface given by $\alpha \rsw$, where $\alpha = 8$.  The fact that the
capture radius is larger than the Schwarzschild radius is due to the
curvature of space by the black hole and is a standard relativistic
effect (see MTW). Using equation [\ref{eq:density}] to specify the
density, we can evaluate the integral to obtain a differential
equation for the time evolution of the black hole mass, i.e., 
\be
{d \mbh \over dt} = {\dot M}_{\rm bh} = {\dot M} (1 - \mu_C) \, , 
\ee
where $\mu_C$ is the cosine of the angle of the last streamline
(measured from the rotational pole) that falls directly onto the
central black hole. Using the orbit equation [\ref{eq:orbit}], we find 
\be
\mu_C = \bigl( 1 - \alpha \rsw/R_C \bigr)^{1/2} \, , 
\ee
where we have evaluated $\mu_C$ at the black hole surface (keep in
mind that $\mu_C = 0$ for $R_C < \rsw$). Next we define dimensionless
variables 
\be
f \equiv \mbh / M_C \qquad {\rm and} \qquad 
\xi \equiv M/M_C \, . 
\ee
The reduced black hole mass $f$ obeys the ordinary 
differential equation 
\be
{df \over d\xi} = 1 - \bigl( 1 - f/\xi^3 \bigr)^{1/2} \, . 
\label{eq:diffeq} 
\ee
Equation [\ref{eq:diffeq}] must be integrated subject to the boundary
condition $f = 1$ at $\xi = 1$ (which means physically that the black
hole mass $\mbh = M_C$ when the centrifugal barrier $R_C$ first
exceeds the Schwarzschild radius $\rsw$, which is true by definition). 
If no additional physical effects prevent continued accretion onto 
the black hole, the enhancement factor is determined by numerically
integrating equation [\ref{eq:diffeq}] to find the limiting value as 
$\xi \to \infty$. In this limit, the black hole mass grows by an 
additional factor $\fa$ as infall continues along the rotational poles 
of the system, where 
\be 
\fa \equiv \lim_{\xi \to 0} f(\xi) \approx 1.3502 \, . 
\label{eq:enhance} 
\ee
In other words, the final black hole mass $\mbh \approx \fa M_C$, 
with $\fa \approx 1.35$, in the absence of additional physical effects. 

However, two additional processes can affect this prediction. First,
the infalling material can experience shell crossings and the baryonic
material can be heated to the system's virial temperature.  As a
result, a hot corona forms around the black hole and the infalling
material must cross this corona in order to become part of this black
hole (anonymous referee; private communication). This effect acts to
reduce the effective value of $\fa$ and hence equation
[\ref{eq:enhance}] provides an upper limit. In other words, the
enhancement factor $\fa$ is confined to the range $1 < \fa < 1.35$.
Second, additional material can be added to the black hole through
disk accretion; this process is addressed in \S 3.1.

We can now evaluate the mass scale for forming black holes using our
fiducial value of the initial rotation rate $\Omega$ (see \S 2.1),
equation [\ref{eq:mcrit}] to set the critical mass scale, and equation
[\ref{eq:enhance}] to specify the enhancement factor $\fa$. We thereby 
find the $\mbh - \sigma$ relation in the form 
\be
\mbh = \fa { 4 \sigma^4 \over G c \Omega} \, \approx 
10^8 \Msun (\sigma/200\kms)^4 \, , 
\label{eq:msig} 
\end{equation}
where we have written the result in terms of $\sigma$ rather than
$\aeff$. This relation is in reasonably good agreement with the 
observed correlations (see equation [\ref{eq:observed}], Figure 1, 
and Paper I).  

Scatter in the value of the initial rotation rate $\Omega$ will
produce corresponding scatter in the resulting $\mbh - \sigma$
relation. To obtain an estimate of this effect, suppose that the
distribution of $\Omega$ has the same form as the distribution of 
the spin parameter $\lambda$ for galactic halos, where numerical 
simulations suggest that $P(\lambda) \propto \exp[ - (\ln
\lambda/\lambda_0)^2/2 \sigma_\lambda^2]$, with $\sigma_\lambda
\approx 0.5$ (Bullock et al. 2001). If the initial rotation rate
$\Omega$ in this model has the same variance, then the $\mbh-\sigma$
relation will develop scatter at the level of $\sigma_\lambda/\ln10$ =
0.22 dex. This level of scatter is represented in Figure 1 by the
bold-faced error bar symbols on the theoretical curve. For
completeness, we note that variations in the enhancement factor $\fa$
will introduce additional scatter into the $\mbh-\sigma$ relation.  If
the parameter $\fa$ is uniformly distributed over the range $1 \le \fa
\le 1.35$, the resulting scatter will be about 0.04 dex.  Since the
combined variances of $\Omega$ and $\fa$ add in quadrature (e.g.,
Richtmyer 1978), the resulting scatter is approximately 0.224 dex 
(safely smaller than the observed scatter of 0.30 dex). 

Most of the baryonic material not captured by the black hole during
this early collapse phase eventually forms stars in the galactic
bulge.  Dark matter with low angular momentum is captured into the
black hole along with the baryons; dark matter with high angular
momentum ($p > 4 \rsw$) passes right through the galactic plane and
forms an extended structure. 

\subsection{Bulge Mass Scale and Mass Ratios} 

This simple dynamical model also predicts a mass scale $M_B$ for the
bulge itself: In the absence of additional physical processes, the
collapse of a structure with initial conditions described by equation
[\ref{eq:initial}] will produce a ``bulge structure'' with a
well-defined mass scale. Unfortunately, bulge formation is complicated
by a host of additional processes (see, e.g., Kauffmann, White, \&
Guiderdoni 1993; Cole et al. 1994; Somerville \& Primack 1999; and
especially Kauffmann \& Haehnelt 200). The baryonic gas must cool to
form the bulge, and the cooling time can be quite long (much longer
than the collapse time scales -- see the following subsection).
Additional gas can be expelled from the bulge through the action of a
galactic wind.  The dark matter can undergo violent relaxation and
only some fraction of the dark matter will remain within the bulge
itself.  Finally, a significant fraction of the baryonic material can
form the inner portion of the galactic disk, rather than remain in the
bulge.  In spite of these complications, it is interesting to find the
mass scale $M_B$ defined by the dynamical model alone:

If the initial protobulge structure is rotating at angular velocity
$\Omega$, then only material within a length scale $R = \aeff/\Omega$
can collapse to form the bulge. Material at larger radii, $r > R$, is
already rotationally supported and will not fall inwards. In the
absence of the aforementioned additional processes, the length scale
$R$ thus defines an effective outer boundary to the collapsing region
that forms the bulge (see Terebey et al. 1984 for a detailed
calculation of how the rotating, collapsing inner region can match
smoothly onto the static, uncollapsing outer region). The boundary
$R$, in turn, defines a mass scale for the bulge, $M_B \sim M(R)$,
i.e., 
\be
M_B = {\cal F}_B \fdm {2 \aeff^3 \over G \Omega} \, \approx \, 
3.3 \times 10^{10} M_\odot \fdm (\sigma / 200 \kms)^3 \, \, , 
\label{eq:mbulge}
\ee
where the second approximate equality assumes $\fdm$ = 1 and 
${\cal F}_B$ $\approx \pi/2$. The factor ${\cal F}_B$ takes into
account the fact that material along the poles can fall into the bulge
even though material in the equatorial plane is rotationally supported.
The value ${\cal F}_B = \pi/2$ is determined by integrating the
initial density distribution (equation [\ref{eq:initial}]) over the
entire cylinder defined by $\varpi < R = \aeff/\Omega$ (where $\varpi$
is the cylindrical radius). The fraction $\fdm \le 1$ is the fraction
of the initial mass that is retained within the bulge structure; for
example, not all of the dark matter necessarily remains in the bulge.

The resulting expression (equation [\ref{eq:mbulge}]) shows that the
bulge mass scale exhibits power-law behavior with $M_B \sim \sigma^3$.
It is interesting to compare this result to the observed bulge masses
in the sample of galaxies with central black holes. Figure 2 shows the
mass scale of equation [\ref{eq:mbulge}] along with the observed data;
also shown is the best fit power-law, which has slope $\sim 3.3 \pm 0.1$. 
For comparison, the traditional Faber-Jackson relation for elliptical 
galaxies and bulges has the form $L \propto \sigma^4$ (Binney \&
Merrifield 1998; Faber \& Jackson 1976).  Consistency of equation
[\ref{eq:mbulge}] with the Faber-Jackson law would require that 
$L \propto M_B^{4/3}$. For a constant stellar IMF, this relation, in
turn, would imply a star formation efficiency $\epsilon$ of the form
$\epsilon \sim M_B^{1/3}$. In other words, in order for the mass scale
of equation [\ref{eq:mbulge}] to describe the masses of observed
bulges, the more massive systems would have to be more effective at
forming stars.

We are also interested in the ratio of black hole mass to bulge mass.
In our simple dynamical model, the bulge mass scale $M_B$ and the
black hole mass $\mbh \approx M_C$ have the same functional dependence
on the rotation rate $\Omega$. The resulting ratio $\mbh/M_B$ $\equiv$
$\mrat$ is independent of $\Omega$ and is given by 
\be 
\mrat \equiv {\mbh \over M_B} = {\sqrt{32} \over {\cal F}_B} 
\, {\sigma \over c} \approx 0.0024 \, (\sigma/200 \kms) \, .  
\label{eq:fraction} 
\ee 
This mass fraction $\mrat$ is roughly comparable to the observed ratio
of black hole masses to bulge masses in host galaxies. The first
estimates suggested that this mass ratio is nearly constant (e.g.,
Richstone et al. 1998; Magorrian et al. 1998), although the data show
appreciable scatter. Later papers found values of $\mrat$ = 0.0015 --
0.0020 (e.g., Ho 1999; Kormendy 2000), in reasonable agreement with
the typical value suggested by equation [\ref{eq:fraction}].  More
recent work (Laor 2001) finds that the mass ratio is an increasing
function of bulge mass, $\mrat \propto$ $M_B^{0.53 \pm 0.14}$, with
$\mrat \approx 0.0005$ for the smallest bulges with observed black
holes and $\mrat \approx 0.005$ for bright ellipticals (cf. McLure \&
Dunlop 2002); this latter result is somewhat steeper than the prediction 
of this model, which implies $\mrat \propto M_B^{1/3}$. The law 
[\ref{eq:fraction}] is shown in Figure 3 along with the observational
data. An unweighted fit to the data implies a slope of $\sim 0.9$
(close to the model prediction of 1.0), but the error bars and scatter
in the data are too large to make a definitive claim. Notice also that 
the observed black hole scaling law, $\mbh \sim \sigma^4$, and the 
observed scaling law for bulge masses, $M_B \sim \sigma^{3.3}$, imply 
that $\mrat \sim \sigma^{0.7}$. 

\subsection{Time Scales} 

With an initial density profile of the form $\rho \sim r^{-2}$, a
detailed collapse calculation (Shu 1977) indicates that the flow
exhibits a well defined mass infall rate $\dot M$ = $m_0 \aeff^3 /G$,
where $m_0 \approx 0.975$ [notice that this starting state corresponds
to an unstable hydrostatic equilibrium]. The infall rate is constant
in time and we can measure the time elapsed since the collapse began
by the total mass $M$ that has fallen to the galactic center. At early
times, all of the mass falling to the center is incorporated into the
central black hole. At later times, the mass supply is abruptly shut
off by conservation of angular momentum. In this setting, the mass
infall rate is quite large, $\dot M$ $\approx$ 650 $M_\odot$ yr$^{-1}$
(for $\sigma$ = 200 \kms and $\sigma^2 = 2 \aeff^2$). The time scale
$\tbh$ to form a typical supermassive black hole (with mass $\mbh \sim
10^8$ $M_\odot$) is about $\tbh \sim 10^5$ yr, comparable to the time
scale $\tstar$ for individual stars to form (e.g., Adams \& Fatuzzo
1996; Myers \& Fuller 1993). In the absence of any competing physical
processes, the dynamical time scale to form the entire bulge is much
longer, about $\tblg \sim 25 - 50$ Myr, comparable to the crossing
time $\tcross$ = $R/\aeff$ of the initial structure. One should keep
in mind that bulge formation also requires the gas to cool, however,
and the cooling time scale can be longer than this dynamical time
scale.

\section{GENERALIZATIONS OF THE MODEL} 

In this section we present further generalizations of this collapse
model for black hole formation. The previous section shows how the
black hole gains mass through infall from the collapse flow. However,
additional mass can be added to the black hole through disk accretion
(\S 3.1).  Furthermore, black holes formed through this collapse
picture are born rapidly rotating (\S 3.2).  The scaling relation for
the black hole mass $\mbh$ as a function of $\sigma$ depends on the
initial angular momentum profile of the pre-collapse material; in
fact, the angular momentum distribution is the most important
determining factor in specifying the black hole masses (see \S 3.3,
3.4). This model can also be generalized include the effects of
mergers (\S 3.5) and non-spherical initial conditions (in particular,
quadrapole moments; see \S 3.6).

\subsection{Disk Accretion} 

Material that falls to the midplane of the system in gaseous form can
collect into a disk structure that surrounds the nascent black hole. 
The presence of the disk is consistent with the current theoretical
ideas about active galactic nuclei and the jets they produce. In order
to retain the desired scaling law $\mbh \sim \sigma^4$, however, the
total amount of mass added to the black hole through disk accretion
should be less than (or at most comparable to) the original mass scale
$M_C$.

The energy density of the universe from quasar activity places a limit
on the amount of mass that can be accreted by black holes. This energy
density $U_T$ has been estimated to be $U_T \approx 2.5 \times
10^{-15}$ erg/cm$^3$ (Elvis, Risaliti, \& Zamorani 2002).  If this
energy arises from mass accretion onto black holes with energy
conversion efficiency $\epsilon$, then the energy density $U_T$
implies a corresponding minimum mass density in black holes at the
present cosmological epoch. Two recent papers (Elvis et al. 2002; Yu
\& Tremaine 2002) have estimated the amount of mass accreted through
quasar activity over the reshift range $0 \le z \le 5$. The mass
density in black holes due to accretion activity takes the form
$\rho_{acc} \approx 2.1 \times 10^5$ $[0.1 (1-\epsilon)/\epsilon]$
$M_\odot$ Mpc$^{-3}$.  For comparison, the observed $\mbh-\sigma$
relation implies a present day mass density in black holes $\rho_{\rm
bh}$ $\approx$ $2.5 \times 10^5 M_\odot$ Mpc$^{-3}$ (Yu \& Tremaine
2002; Aller \& Richstone 2002). The present ratio of the accreted mass
to the observed mass is thus $\cal R$ = 0.083 $(1-\epsilon)/\epsilon$.
In order for the infall collapse model of this paper to explain the
observed black hole masses, the ratio ${\cal R} < 1$ and hence the
conversion efficiency must be at least $\epsilon \sim 0.08$. If the
efficiency of energy conversion in quasars is too low, then black
holes would gain more mass from disk accretion than from infall. If
the energy conversion efficiency is high, however, quasar light could
be explained by a modest addition of mass, and some other physical
process (e.g., infall) would be required to explain the observed black
hole masses.

Because of the present observational uncertainties, the critical value
of $\epsilon_C$ (the value required for disk accretion to explain the
observed mass density in black holes) is not completely specified.
Using only the quasar constraint, Yu \& Tremaine (2002) find
$\epsilon_C \approx 0.1$, whereas Elvis et al. (2002) find $\epsilon_C
\approx 0.15$.  For Schwarzschild black holes, the energy conversion
efficiency is expected to be about $\epsilon \sim 0.1$, but higher
efficiency, with $\epsilon \sim 0.2$, is possible for thin-disk
accretion onto a Kerr black hole (see Yu \& Treamine 2002). A
plausible upper limit for accretion processes is $\epsilon = 0.31$
(Thorne 1974). Both the observations of background radiation and the
theoretical expectations for energy conversion efficiency $\epsilon$
should be specified further to resolve this issue. 

For comparison, we can derive another constraint on the amount of
accreted mass. This constraint is more general than in the discussion
above, but is also weaker. We consider the limiting case in which disk
accretion is maximally effective. Disk accretion generally cannot
operate faster than the orbit time at the outer disk edge (Shu 1992). 
In this context, the orbit time $\tau$ is given by 
\be
\tau^2 = {4 \pi^2 R_C^3 \over GM} = 
{4 \pi^2 \Omega^6 (GM)^8 \over \sigma^{24} } \, . 
\label{eq:tau} 
\ee
When the disk accretion time becomes longer than the time required for
the disk to condense into stars, i.e., when $\tau > \tstar$, disk
accretion is no longer effective and the mass flow onto the central
black hole must come to an end. This condition implies a maximum mass
scale for accreting black holes. The mass appearing in equation
[\ref{eq:tau}] above represents the total mass that has fallen to the
center by a given time. Only a fraction of this mass is available to
join the disk because only a fraction $f_B$ is baryonic and a fraction
$f_G$ is in gas (rather than in stars).  Including these two factors,
the maximum mass that can be added to the black hole via disk
accretion becomes 
\be
M_{\rm max} = (2 \pi)^{-1/4} f_B f_G (\sigma^3/G) 
(\tstar/\Omega^3)^{1/4} \, . 
\ee
For typical values of the input parameters $\sigma$ and $\Omega$, and
for $f_B$ = 2/15, $f_G$ = 1/2, this maximum mass scale is about 5
times larger than $M_C$.  In the limit of maximally efficient disk
accretion, the mass scale $M_C$ defined by the centrifugal radius can
thus be compromised (see also Burkert \& Silk 2001; Silk \& Rees
1998). Unfortunately, the disk accretion rates are not known in these
early stages of galaxy formation (however, see Kumar 1999). In most
known astrophysical disks, the disk accretion rates are much smaller
than their maximum values, typically by factors of $10^2 - 10^4$
(e.g., Shu 1992), which would imply that most of the black hole mass
does not come from accretion. On the other hand, as discussed above, 
the observed X-ray background implies that the central black holes
that power quasars must accrete a substantial mass, roughly comparable 
to the masses obtained via infall. This argument assumes that the mass 
in the disk is large enough to affect the black hole mass; however, the 
infall model used here naturally places most of the incoming mass at 
large radii appropriate for the disk (e.g., Cassen \& Moosman 1981). 

\subsection{Black Hole Angular Momentum} 

This theoretical model for supermassive black hole formation predicts
that the resulting black holes should be rotating rather rapidly. If
we approximate the orbital solutions for incoming material using the
classical treatment of \S 2.2 and the effective capture radius of a
Schwarzschild black hole, then the angular momentum $\jbh$ of the
resulting black hole takes the form 
\be
\jbh = {1 \over 9} \, 2^{11} \, \aeff^8 \, 
c^{-3} \, G^{-1} \, \Omega^{-2} \, . 
\ee
In the absence of additional mass sources, we can write this angular 
momentum in terms of the black hole mass $\mbh = M_C$ and the 
Schwarzschild radius $R_S$, 
\be
\jbh = {4 \over 9} {c \mbh \rsw} \, \approx 0.44 c \mbh \rsw \, . 
\label{eq:jbhole} 
\ee 
The maximum allowed value for the numerical coefficient in equation
[\ref{eq:jbhole}] is 0.5 (e.g., Thorne, Price, \& MacDonaold 1986;
Blandford 1990), so these black holes are rotating close to their
maximum rates. Notice that relativists usually write the Kerr metric
in terms of the parameter $a \equiv J/M$. In units where $G=1$, this
parameter has a maximum value of $a = M$. This theory predicts the
formation of black holes with the ratio $a/M$ $\approx$ 0.89 -- close
to its maximum value of unity. Continued infall will add both mass and
angular momentum to the black hole, and will change this prediction
somewhat. Notice also that disk accretion models predict that
supermassive black holes should be rotating rapidly (e.g., Elvis et
al. 2002).

Keep in mind that this result uses mixed approximations. We have
derived the maximum mass scale using the capture radius appropriate
for Schwarzschild geometry and purely classical orbits solutions.  The
resulting black holes are rapidly rotating, however, and curve
space-time as described by the Kerr metric. A fully self-consistent
calculation should find orbit solutions and the capture radius for
Kerr black holes, with the angular momentum parameter $a_J =
\jbh/\mbh$ determined at each evolutionary stage by conservation of
angular momentum. This calculation is beyond the scope of this present
paper, but could change our results at the level of 50 percent. 

The last stable orbit for the Kerr metric can be worked out in terms
of simple functions (e.g., Riffert 2000); the last stable orbit for
Kerr geometry is smaller than that of Schwarzschild geometry, and the
last captured streamline should behave similarly. In other words, a
Kerr black hole has a smaller surface area than a Schwarzschild black
hole of the same mass (e.g., Rees 1984). As the black hole grows, it
gains both mass -- which makes its capture cross section larger -- and
angular momentum -- which makes its capture cross section
smaller. Near the end of the infall phase, the added angular momentum
to the black hole decreases its cross section, and the infalling
material starts to have too much angular momentum to fall within the
older, larger cross section. This effect thus acts to make the
transition more abrupt. 

\subsection{General Initial Conditions} 

The success of this simple model for black hole formation rests on the
initial angular momentum profile, which must have a particular
form. In order to determine how sensitive the results are to the
assumed initial conditions, we consider generalized initial density
distributions of the form 
\be
\rho(r) = A r^{-\Gamma} \, , 
\ee
where $A$ is a constant and $\Gamma$ is an arbitrary index. As before, 
we assume that the initial structure is rotating at a uniform rate 
$\Omega$. The corresponding mass distribution is given by  
\be
M(r) = {4 \pi A \over 3-\Gamma} r^{3-\Gamma} \equiv \acon 
r^{3-\Gamma} \, , 
\ee
where we have defined a reduced constant $\acon \equiv 4 \pi
A/(3-\Gamma)$.  To find the black hole mass scale resulting from the
collapse of this initial state, we use the same criterion as before,
which takes the form $j_\infty = 4 GM/c$, where $j_\infty = \Omega$ 
$r_\infty^2$. To specify the starting radius $r_\infty$, we must 
invert the mass distribution to obtain
\be
r_\infty = \Bigl[ {M \over \acon} \Bigr]^{1/(3-\Gamma)} \, . 
\ee 
Solving for the black hole mass scale, we obtain 
\be
\mbh = \Bigl[ 
{4 G \over c \Omega} \Bigr]^{(3-\Gamma)/(\Gamma-1)} 
\acon^{2/(\Gamma-1)} \, , 
\ee
where we must restrict the analysis to $\Gamma > 1$. For less steep
initial density distributions, the centrifugal barrier of the collapse
flow increases more slowly with mass than does the capture radius of
the black hole (which has a linear dependence).

We can also solve for the other parameters of the forming 
bulge system. The radius $R$ that marks the outer boundary is 
given by 
\be
R = \Bigl[ {\acon G \over \Omega^2} \Bigr]^{1/\Gamma} \, , 
\ee
while the bulge mass scale takes the form 
\be
M_B = 
\Bigl[ {G \over \Omega^2} \Bigr]^{3/\Gamma-1} 
\acon^{3/\Gamma} \, . 
\ee
With the radius $R$ and bulge mass scale defined, we can solve 
for the expected velocity dispersion of the final system using 
the relation $\sigma^2 \approx 2 G M_B / R$, where the factor of 
2 arises from the collapse itself. The resulting expression for the 
velocity dispersion is 
\be
\sigma = \sqrt{2} \, \Omega^{1 - 2/\Gamma} \, 
\Bigl[ \acon G \Bigr]^{1/\Gamma} \, .
\ee
With this expression for $\sigma$ in hand, we can write the expected
bulge mass $M_B$, bulge size scale $R$, and black hole mass $\mbh$ in
terms of $\sigma$ rather than the initial variable $A$ (or $\acon$)
appearing in the density distribution, i.e., 
\be 
M_B = {\sigma^3 \over 2 \sqrt{2} \Omega G} \, , \qquad 
R = \sigma / (\sqrt{2} \Omega) \, , \qquad {\rm and} \qquad 
\mbh = {\sigma^4 \over c G \Omega} \Bigl[ \sqrt{8} 
{\sigma \over c} \Bigr]^{(4-2\Gamma)/(\Gamma-1)} \, . 
\ee 
And finally, we obtain the corresponding expression for the mass 
ratio: 
\be
\mrat = \Bigl[ \sqrt{8} 
{\sigma \over c} \Bigr]^{(3-\Gamma)/(\Gamma-1)} \, . 
\ee
This result is {\it very} sensitive to the starting slope $\Gamma$ 
of the density profile (for an assumed constant rotation rate 
$\Omega$). If the index $\Gamma$ is much smaller (larger) than the
preferred value $\Gamma = 2$, then the final black hole masses are
much smaller (larger) than those observed. Variations in the slope 
$\Gamma$ will produce corresponding scatter in the observed 
$\mbh-\sigma$ relation; as a benchmark, variations at the level of 
$\Gamma = 2 \pm 0.1$, will produce scatter of about 0.5 dex. The 
value of the index $\Gamma$ affects this model for mass scales $M_S$ 
in the range $0.1 \mbh < M_S < 0.5 M_B$. 

In summary, this model is sensitive to the form of the initial
conditions. Moderate departures from our assumed starting condition
$\rho \sim r^{-2}$ to large variations in the predicted black hole
masses $\mbh$ and mass ratios $\mrat$. If this model is correct, then
the relevant initial conditions for the collapse of galactic bulges
{\it must} have angular momentum distributions close to those assumed
here (specified by equation [\ref{eq:initial}]).

\subsection{Dimensional Analysis} 

Given all of the generalized relations discussed in the previous
section, how can we make sense of all the possibilities? For this
theoretical model, the story is relatively simple: For the formation
of the bulge itself, the black hole forming at the galactic center has
essentially no effect. Because no relativistic effects come into play,
the speed of light $c$ does not enter into the formulae describing the
bulge properties $\sigma$, $M_B$, and $R$. The only variables that can
determine these quantities are the rotation rate $\Omega$, the
parameters of the initial density distribution ($A$ and $\Gamma$), and
the gravitational constant $G$. Furthermore, the index $\Gamma$ is
dimensionless, so the only quantities that carry dimensions are $A$,
$\Omega$, and $G$. A simple virial argument lets us replace the
density coefficient $A$ with the velocity dispersion $\sigma$, which
is a much more familiar quantity. For a given velocity scale $\sigma$,
the quantity $\sigma^3/G$ defines a mass infall rate; when combined
with a time scale $\Omega^{-1}$ (which is the only time scale present
in this simple treatment) we thus obtain the mass scale $M \sim
\sigma^3/G \Omega$, which we identify as the bulge mass scale $M_B$.
To summarize, the three quantities with dimensions ($A$, $G$,
$\Omega$) can only define the three bulge quantities ($\sigma$, $M_B$,
$R$) in one way (up to dimensionless factors of order unity).

Next, however, we must put the black hole mass into the problem.  
The black hole introduces relativistic effects and, in particular,
introduces the speed of light $c$ as another fundamental constant.
The ratio $\beta = \sigma/c$ is another dimensionless parameter in the
problem. In terms of dimensional analysis, we can now define an
infinite number of new scales. For example, in addition to the bulge
mass scale $M_B$, we now have the family of mass scales $M = F(\beta)
M_B$, where $F(\beta)$ is an unspecified function of the second
dimensionless field $\beta$. In our treatment, the power-law forms for
the initial conditions lead to new mass scales given by the power-laws
$M_\eta = M_B \, \beta^\eta$, where $\eta$ can be any real number.
The physical law of angular momentum conservation, in conjunction with
the initial profile, chooses the value of the exponent $\eta$ for a
given model. In the simplest case, that of isothermal $\rho \sim
r^{-2}$ initial conditions, we obtain $\eta = 1$ and hence $\mbh \sim
M_B (\sigma/c)$, which happens to be the observed scaling law ($\mbh
\sim \sigma^4$). Notice also that the ratio $\mbh/M_B \sim \sigma/c$
$\sim 10^{-3}$ (the observed order of magnitude).  The other (more
general) choices of initial conditions, $\rho \sim r^{-\Gamma}$, thus
correspond to other possible values of the exponent $\eta$, i.e.,
$\eta$ = $(3-\Gamma)/(\Gamma-1)$.

\subsection{Survival of Scaling Laws with Mergers} 

Many galaxies are expected to experience merger events during their
formative stages of evolution (e.g., White \& Rees 1978; White 1979).
If the initial collapse of protobulges proceeds as envisioned in the
simple theoretical model developed in this paper, then the fundamental
building blocks of galaxies will have bulge mass $M_B \sim \sigma^3$
and black hole masses $\mbh \sim \sigma^4$.  However, observations
indicate that the {\it final} merger products obey these scaling
relations. As a result, we must now consider what happens to these
scaling laws if the fundamental building blocks undergo multiple
merger events (see also Hughes \& Blandford 2002; Menou, Haiman, \&
Narayanan 2001; Ebisuzaki et al. 2001; Volonteri, Haardt, \& Madau
2003).

For the sake of this discussion, we consider the rotation rate to be
constant so that the only relevant variable is the velocity dispersion
$\sigma$. We define a dimensionless variable $s = \sigma/(200
\kms)$. If protobulges merge many times and if they all contain central
black holes that merge, then the final black hole mass can be written
in the form 
\be
\mbh_{f} = \sum_j \mbh_j = M_0 \sum_j s_j^4 \, , 
\label{eq:massadd} 
\ee
where we assume in the second equality that the starting units,
labeled by the index $j$, obey the scaling law derived in \S 2 (and
Paper I). All sums are taken up to $N$, which specifies the number of
protobulges (initial units) that merge to form the final stellar
system. Notice that this equation assumes minimal energy losses 
from gravitational radiation during the collisions. 

We must relate the velocity dispersion $s_f$ of the final bulge system
to the velocity dispersions $s_j$ of the individual units.  We first
consider one idealized limiting case: If the mergers take place with
zero orbital energy and no energy losses occur during the collision,
then the final energy of a merged system must equal the internal
energies of the initial (pre-merger) pairs.  If we consider the
systems to be in virial equilibrium and to be homologous (a gross
approximation, but a good place to start for conceptual purposes),
we obtain the relation 
\be
s_f^2 = \sum_j p_j s_j^2 \, , 
\label{eq:sigadd} 
\ee
where we have defined $p_j \equiv R_{Bj}/R_{Bf}$. 

Now let us define $q_j \equiv s_j^2$ for all protobulges labeled 
by the index $j$. We can think of the $q_j$ and the $p_j$ as 
components of $N$-dimensional vectors (which live in the space 
of protobulge parameters). With this ansatz, we can write the 
final black hole mass in the form 
\be
\mbh_f = M_0 \sum_j q_j^2 = M_0 |q_j|^2  \, , 
\ee
where the notation $|q_j|^2$ denotes the vector magnitude squared; 
the final velocity dispersion $s_f$ takes the form 
\be
s_f^2 = {\bf p} \cdot {\bf q} \qquad \Rightarrow \qquad 
s_f^4 = |p_j|^2 |q_j|^2 \cos^2\thb \, , 
\ee
where $\thb$ is the angle between the two vectors in bulge
parameter space. Combining the above equations, the final black 
hole mass takes the suggestive form 
\be
\mbh_f = {M_0 s_f^4 \over |p_j|^2 \cos^2\thb} \, . 
\ee 
The final mass of the black hole thus displays a quartic dependence 
on the final velocity dispersion (to leading order). Some intrinsic 
scatter will be introduced through varying angles $\thb$ and mass
weights $p_j$. We can consider the limiting case in which all of the
starting protobulge units are identical, $p_j = 1/N$ for all units
$j$, and the two vectors $p_j$ and $q_j$ are parallel so that $\cos
\thb = 1$. In this limit, the final black hole mass takes the form
$\mbh = N M_0 s_f^4$, which is consistent with the observed scaling
relation (equation [\ref{eq:observed}]). In other words, the scaling
law can be preserved under the action of mergers in this idealized
limit.

We can also find the behavior of the bulge mass under repeated 
merger events. The final bulge mass $M_{Bf}$ after $N$ mergers is 
given by 
\be
M_{Bf} = M_{B0} \sum_j s_j^3 = M_{B0} \sum_j s_j q_j 
= M_{B0} \, {\bf s} \cdot {\bf q} \, . 
\ee
We can write the dot product in terms of another angle $\thbulg$, 
\be
{\bf s} \cdot {\bf q}  = |s_j| \, |q_j| \, \cos \thbulg \, , 
\ee
which allows us to write the final bulge mass in the form 
\be
M_{Bf} = M_{B0} |s_j| s_f^2 \Bigl\{ 
{\cos\thbulg \over |p_j| \cos\thb} \Bigr\} \, . 
\ee 

To explicitly illustrate the possibility of preserving the $\mbh -
\sigma$ relation under the action of mergers, we have performed a
simple set of Monte Carlo simulations, as depicted in Figure 4. In
these numerical experiments, the number of interacting units is
randomly chosen within the range 2 $\le N \le$ 6 (see, e.g., Wechsler
et al. 2002).  The velocity dispersions of the interacting units are
chosen to be random, but logarithmically spaced in order to evenly
sample the observed range of (final) $\sigma$. The radial sizes of the
interacting units $p_j = R_{jB}/R_{fB}$ are chosen to be equal so that
$p_j = 1/N$. In order to get the normalization correct, we let the
mass scale $M_0$ of the interacting units (equation
[\ref{eq:massadd}]) be a factor of three smaller than that of the
observationally determined distribution (equation
[\ref{eq:observed}]). Notice that this approximation is equivalent to
allowing serious energy losses during the merger events. The final
values of the black hole mass and velocity dispersion are then
calculated according to equations [\ref{eq:massadd}] and
[\ref{eq:sigadd}]. With these idealizations, the final distribution
retains its power-law form and appears to be in good agreement with
the observed correlation (see Figure 4). We stress, however, that the
range of parameter space available for merging protobulges is enormous
-- not every merger scenario will produce such a clean correlation. We
leave a more detailed exploration of parameter space for future work.

\subsection{\bf The Effects of Initial Quadrapole Moments} 

In this subsection, we consider the effects of a quadrapole moment on
this collapse picture of black hole formation.  Since the protobulge
structures can obtain their initial rotation rates through tidal
torques acting on nonzero quadrapole moments of the mass distribution,
it is important to check whether such quadrapole moments affect the
collapse flow.  We can consider two limiting cases: A substantial
quadrapole moment in the inner regime and a substantial quadrapole on
the large size scale of the protobulge itself. The case of the inner
regime is very much like the binary star potential in the star
formation problem, and this situation has already been shown to have
little effect (Allen 1999, Jijina 1999). We thus consider the case of
an outside quadrapole moment.

As a starting point, we consider the outer quadrapole to be akin to
two point masses $M$ with separation $R$. In order for the outer
quadrapole to exert a torque on the inner region (the portion of the
pre-collapse structure that will eventually comprise the black hole),
the inner region must also have a quadrapole moment. We consider the
inner portion to have a dumbbell shape with mass scale $m$ and size
$\ell$.  The force exerted on the inner region by the outer quadrapole
is given by the tidal force law 
\be
F \approx {G M m \over R^2} {\ell \over R} \, . 
\ee
The torque $\tau$ exerted on the inner region is given by this force 
acting through a lever arm of size $\ell$. The torque is given by 
\be
\tau \approx {G M m \ell^2 \over R^3} \, . 
\ee
Now we need to make the connection between these quantities and the
scales in the black hole formation problem. The size scale $R$ is the
size of the protobulge, i.e., $R \approx \aeff / \Omega$. Because the
mass distribution has the profile of an isothermal sphere ($\rho \sim
r^{-2}$ and $M(r) \sim r$), the size $\ell$ is smaller than $R$ by the
ratio of the black hole mass to the bulge mass, i.e., $\ell/R$ =
$\mu_B$.  The mass $M$ of the outer quadrapole is given by 
$M = \lambda_1 M_B$, where the dimensionless parameter $\lambda_1$ 
determines the asphericity of the configuration. If the distribution 
is perfectly spherical, $\lambda_1 \to 0$. In the limit that the 
protobulge has a dumbbell shape, $\lambda_1 \to 1$. Similarly, we 
let $m = \lambda_2 \mbh$. With these definitions, the torque can 
be written in terms of the physical quantities in our paper as 
\be
\tau = \lambda_1 \lambda_2 \mu_B^2 {G M_B \mbh \over R} \, . 
\ee

One way to quantify the effect of this torque on the collapse flow 
is to find the total change in angular momentum produced by the 
torque, which acts over the collapse time $\Delta t$ of the inner 
region only. (After the black hole has formed, the still-existing 
outer quadrapole will not effect the black hole mass). The time 
for the inner region to collapse is given by 
\be
\Delta t \approx \mbh / {\dot M} = {G \mbh \over \aeff^3} = 
{2 \mbh \over M_B \Omega} = {2 \mu_B \over \Omega} \, . 
\ee 

Combining the above expressions for the torque and the time interval 
over which it acts, we find the change in angular momentum:  
\be
\Delta J = 
2 \lambda_1 \lambda_2 \mu_B^2 {G \mbh^2 \over R \Omega} \, . 
\ee
For later convenience, we note that our model implies $G M_B /R^3$ =
$2 \Omega^2$, so we rewrite the above expression to obtain the form 
\be
\Delta J = 4 \lambda_1 \lambda_2 \mu_B \mbh \Omega \ell^2 \, . 
\ee

The starting angular momentum $J_0$ of the inner region can be 
obtained by integrating over the initial density distribution. 
The result is 
\be
J_0 = {2 \over 9} \mbh \Omega \ell^2 \, . 
\ee
The resulting fractional change in the angular momentum of the
inner result is immediately found to be 
\be
{\Delta J \over J_0} = 18 \lambda_1 \lambda_2 \mu_B \, . 
\ee
The mass fraction $\mu_B \approx 0.0024$ in our model, so this 
fraction change becomes 
\be
{\Delta J \over J_0} \approx 0.043 \lambda_1 \lambda_2 \, . 
\ee
Even in the extreme limit of highly developed quadrapole moments on
both the inside and the outside, the fractional change in angular
momentum is only about 4 percent. In a more realistic case, the outer
quadrapole should be smaller than unity $\lambda_1 < 1$, but still
large enough to give the bulge an elliptical appearance; the inner
region can be smoothed out (e.g., see Peebles 1993 for streaming
arguments) and the inner quadrapole should also have $\lambda_2 < 1$. 
In any case, the coupling between the outer quadrapole moment and the
inner collapse region can be considered weak in the context of our
orbit solutions.

\section{CONCLUSIONS} 

In this contribution, we have presented further development of the
simple theoretical model for supermassive black hole formation that
was put forth in Paper I.  This model begins with an initial state
specified by a density distribution of the form $\rho = \aeff^2 / 2
\pi G r^2$ and a uniform rotation rate $\Omega$.  The parameters
$(\aeff, \Omega)$ represent the specification of the initial
conditions.  As the initial state collapses to form a galactic bulge,
the collapse flow produces a black hole in the center.  The velocity
dispersion of the final bulge system is comparable to the effective
transport speed and we make the identification $\sigma \approx
\sqrt{2} \aeff$. In developing this basic picture, we find the
following results:
 
[1] The black hole mass $\mbh$ is determined by the condition that the
centrifugal radius exceeds the capture radius of the central black
hole. This requirement leads to the scaling law $\mbh = M_0
(\sigma/200 \kms)^4$, which is consistent with observations both in
its dependence on velocity dispersion $\sigma$ and for the mass scale
($M_0 \approx 10^8 \Msun$) of the leading coefficient (see equation
[\ref{eq:msig}] and Figure 1).  The mass scale $M_0$ is given by $M_0$
= $4 \fa (200 {\rm km/s})^4 / c G \Omega$, so that variations in the
rotation rate $\Omega$ and the amount of continued infall $\fa$ lead
to dispersion about the observed power-law correlation. If the initial
rotation rate $\Omega$ follows the same distribution as that
calculated for the spin parameter $\lambda$ of dark matter halos, then
variations in $\Omega$ would produce a scatter of $\sim 0.22$ dex (a
factor of $\sim1.7$) in the predicted $\mbh-\sigma$ scaling law. The
observed relation has a factor of 2 dispersion.
 
[2] A bulge mass scale is defined in this model by the outer boundary
of the collapsing region. Material at initial radii $r > R$ is
rotationally supported and cannot collapse, where $R = \aeff/\Omega$.
This condition implies a scaling law for the bulge mass scale, $M_B =
2 {\cal F}_B \aeff^3 / G \Omega$ $\propto \sigma^3$ (see equation
[\ref{eq:mbulge}] and Figure 2). Although bulge formation must involve
physical processes that are not included in this dynamical mdoel
(e.g., gas cooling, feedback from galactic winds, disk formation),
this scaling law for $M_B$ is nonetheless in good agreement with the
observed relation for host galaxies that contain supermassive black
holes, as shown in Figure 2.

[3] If we interpret the bulge mass scale $M_B$ (see [2] above) as the
bulge mass itself, then this model predicts the ratio $\mrat$ of black
hole mass to bulge mass (equation [\ref{eq:fraction}] and Figure 3).
The theoretical mass ratio is proportional to the velocity dispersion
and has the form $\mrat \approx 0.0024$ ($\sigma$/200 km s$^{-1}$),
roughly comparable to observed mass ratios. A mass ratio $\mrat$ that
increases with velocity dispersion $\sigma$ is consistent with (and
even indicated by) a recent analysis of the observational data (Laor
2001); an unweighted least squares fit to the data shown in Figure 3
implies a slope of $\sim0.9$. We also note that a constant mass ratio
$\mrat$ may be inconsistent with the data: Since the black hole mass
scales as $\mbh \sim \sigma^4$ (Tremaine et al. 2002), a constant mass
ratio $\mrat$ would require the bulge mass to also scale as $M_B \sim
\sigma^4$. However, the observed bulge masses (for systems with
detected black holes) do not indicate such a steep dependence on
$\sigma$; the data suggest an index of approximately 3.3 $\pm$ 0.1
(see Figure 2).

[4] In this scenario, the black hole forms quickly, with a typical 
formation time of $\sim 10^5$ yr. 

[5] The black holes formed through this mechanism are born with rapid
rotation rates. As a result, the geometry in the central regions is
best described by the Kerr metric. Specifically, this model predicts
that supermassive black holes are formed with an initial rotation
parameter $a/M \approx 0.9$, relatively close to the maximum value of
$a/M$ = 1. Such high black hole rotation rates may be detectable by
LISA (the Laser Interferometer Space Antenna).

[6] Although the supermassive black holes produced by this process are
intrinsically relativistic objects, relativistic effects play only a
modest role in the collapse flows that produce them. The most
important effect is that the capture radius of a black hole is larger
than the Schwarzschild radius by a factor of 4 and this effect leads
to black hole masses that are larger by this same factor.

[7] The predicted black hole masses are very sensitive to the initial
conditions. For an initial density distribution of the form $\rho \sim
r^{-2}$ (with constant angular velocity $\Omega$), the subsequent
collapse produces black holes and galactic bulges with the correct
masses and the correct dependence on velocity dispersion $\sigma$.
The correct mass normalization depends on the choice of rotation rate
$\Omega \approx 2 \times 10^{-15}$ rad/s.  Initial density profiles
with shallower slopes, $\rho \sim r^{-\Gamma}$ with $\Gamma < 2$,
produce {\it smaller} black holes with a {\it steeper} slope in the
$\log \mbh - \log \sigma$ plane. In general, the initial profile of
specific angular momentum -- given by the combination of $\rho(r)$ and
$\Omega(r)$ -- determines the final mass scales. This sensitivity on
initial conditions is both the strongest and weakest aspect of the
model: If we can unambiguously determine the angular momentum
distribution of the initial states, we can directly verify or falsify
this theoretical scenario for black hole and bulge formation. We also 
note that our initial conditions apply on (initial) radial scales of 
several kpc. The manner in which these initial conditions match onto 
the Hubble flow and the larger scale structure of the galactic halo 
(at radial scales of many hundred kpc) remain to be determined. 

[8] If galactic bulges and larger galactic structures are formed
through the mergers of smaller constituent pieces, this scenario for
black hole formation can still play a role: In this case, a number of
the constituent pieces would form black holes in their centers through
this mechanism. The resulting scaling laws (e.g., $\mbh \sim
\sigma^4$) can be preserved, or nearly preserved, under mergers for
idealized circumstances (\S 3.5). On the other hand, mergers tend to
reduce the angular momentum per unit mass, so that merger scenarios
predict a lower angular momentum for the resulting black holes (e.g.,
Hughes \& Blandford 2002).  Future measurements of the angular
momentum of supermassive black holes are thus crucial for
discriminating between various formation scenarios.

\acknowledgments
 
We would like to thank Gus Evrard, Karl Gebhardt, and Risa Wechsler
for useful discussions; we especially thank Rosie Wyse for
enlightening discussions regarding bulge rotation rates. Finally, we
thank an anonymous referee for many suggestions that improved the
paper. This work was supported by NASA through the Long Term Space
Astrophysics program and the Space Telescope Science Institute, and by
the University of Michigan through the Michigan Center for Theoretical
Physics.

\newpage 
\begin{figure}
\figurenum{1}
\epsscale{1.0}
\plotone{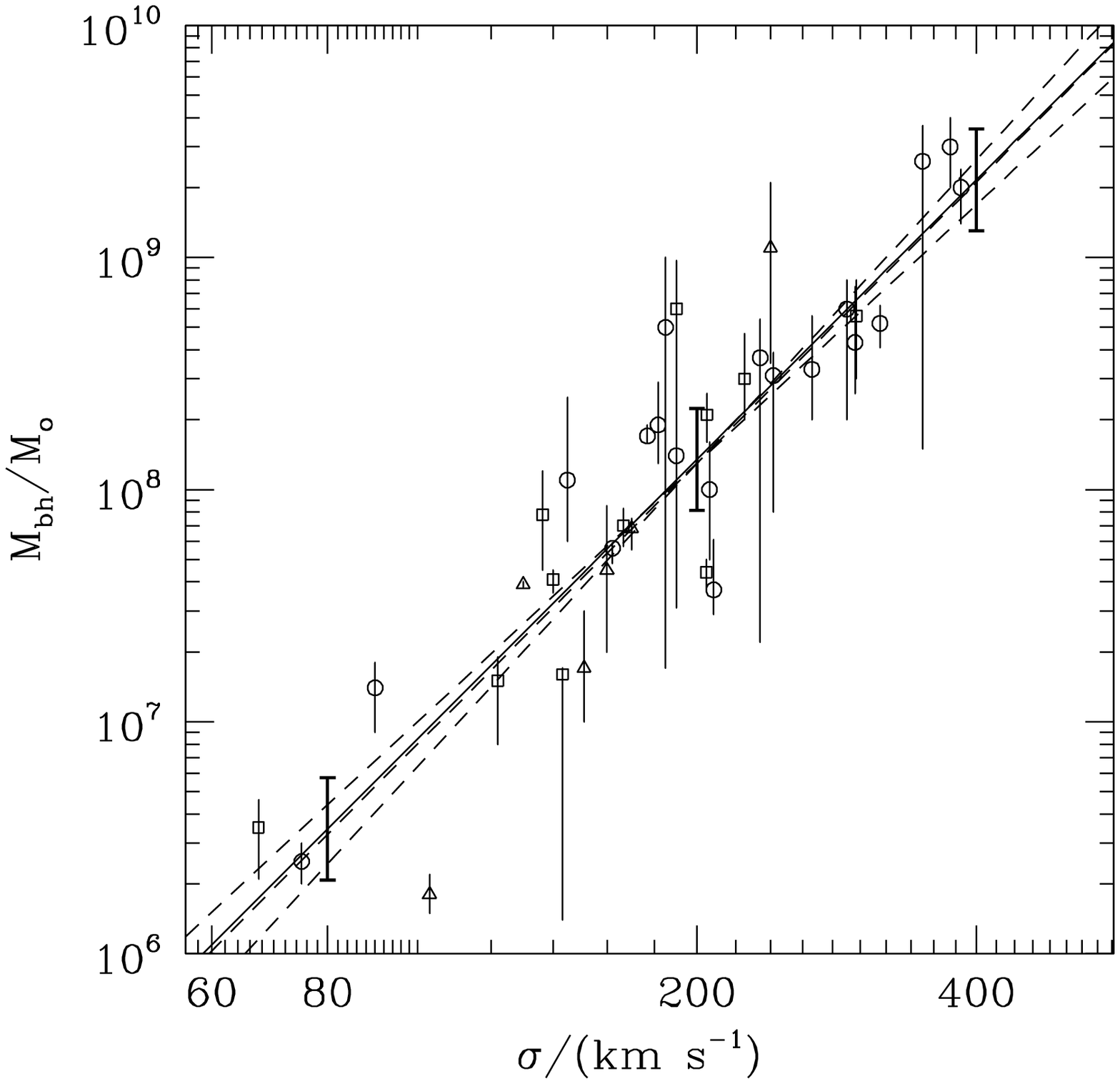}
\figcaption{The correlation between black hole mass $\mbh$ and
velocity dispersion $\sigma$ of the host galaxy. The data points
(adapted from Gebhardt et al. 2003) represent the observed correlation
for ellipticals (circles), S0 galaxies (squares), and spirals
(triangles). The solid curve shows the theoretical result of this
paper (using equation [\ref{eq:msig}] with ${\cal F}_A$ = 1.35). The
dashed curves show the observational fit advocated by Tremaine et
al. (2002); curves are shown for the best estimate of the index
$\gamma$ = 4.02 and for the maximum/minmum values $\gamma$ = 4.02
$\pm$ 0.32. The bold-faced error bar symbols show the level of scatter 
that would result if the initial rotation rate $\Omega$ follows the same
distribution as that of the spin parameter $\lambda$ for galactic
halos.}  
\end{figure}

\newpage 
\begin{figure}
\figurenum{2}
\epsscale{1.0}
\plotone{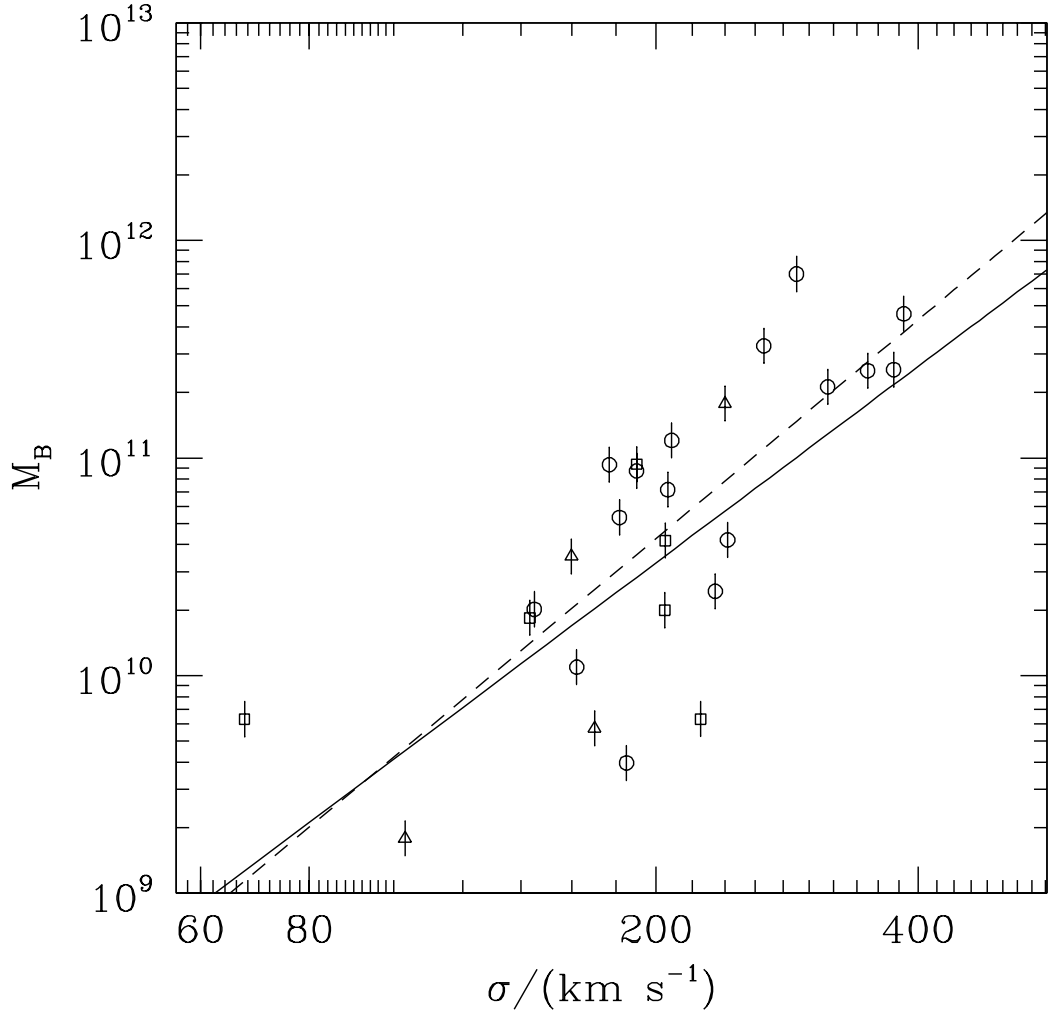}
\figcaption{The correlation between bulge mass scale $M_B$ and
velocity dispersion $\sigma$ of the host galaxy. The data points
(adapted from Gebhardt et al. 2003) represent the observed correlation
for ellipticals (circles), S0 galaxies (squares), and spirals
(triangles).  The error bars correspond to 20 percent uncertainties in
the bulge mass estimates.  The solid curve shows the theoretical mass
scale predicted by the infall-collapse model of this paper (using
equation [\ref{eq:mbulge}] with ${\cal F}_B = \pi/2$ and $\fdm$ = 1).
The dashed curve shows an unweighted least squares fit to the data. }
\end{figure}

\newpage 
\begin{figure}
\figurenum{3}
\epsscale{1.0}
\plotone{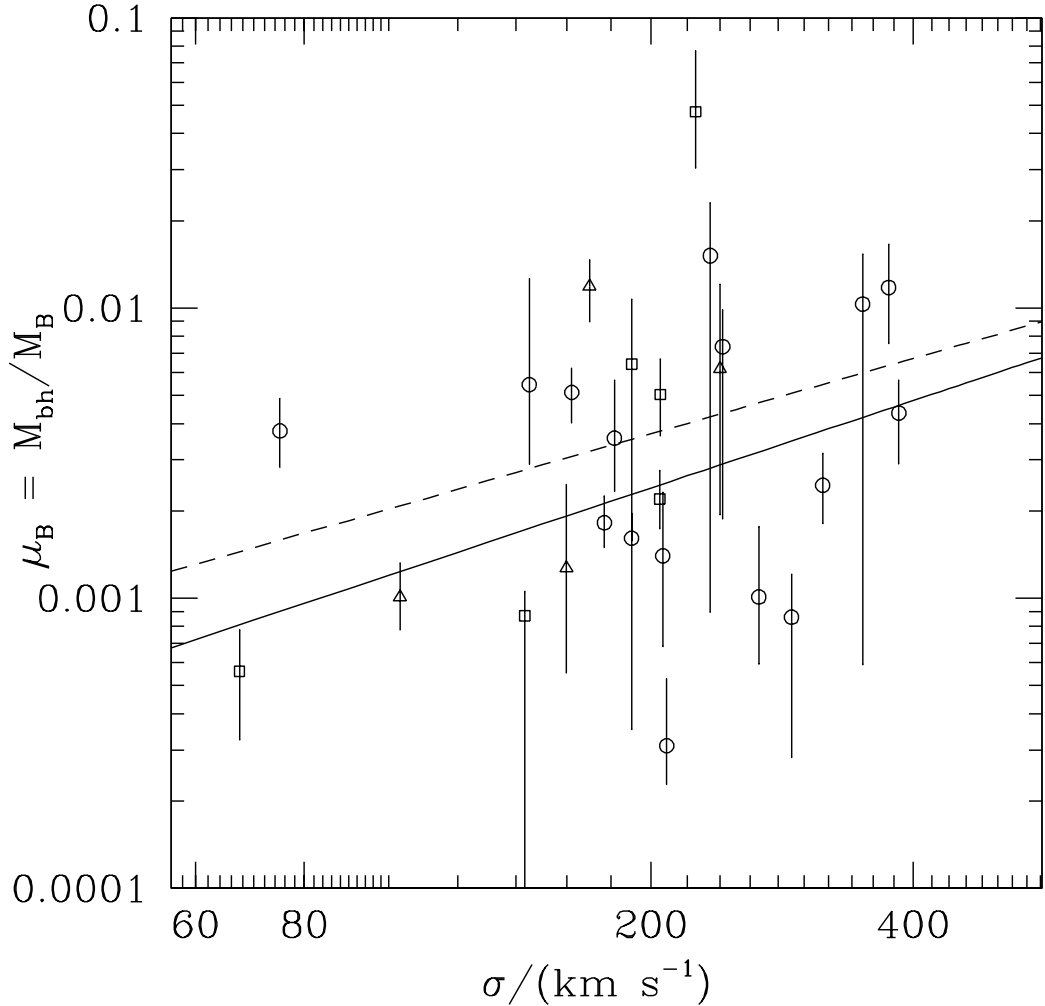}
\figcaption{The ratio $\mrat$ of black hole mass $\mbh$ to bulge mass
scale $M_B$ plotted as a function of the velocity dispersion $\sigma$
of the host galaxy. The solid curve shows the prediction of equation
[\ref{eq:fraction}], where we assume that the simple dynamical mass
scale $M_B$ from the collapse model can be identified with the bulge
mass. The data points (adapted from Gebhardt et al. 2003) exhibit
considerable scatter, but the least squares fit (shown as by the
dashed curve with slope 0.86) is in reasonable agreement with
theoretical expectations; the various symbols represent ellipticals
(circles), S0 galaxies (squares), and spirals (triangles). The error
bars are determined by the quadrature sum of the error bars shown in
Figures 1 and 2. }  
\end{figure}

\newpage 
\begin{figure}
\figurenum{4}
\epsscale{1.0}
\plotone{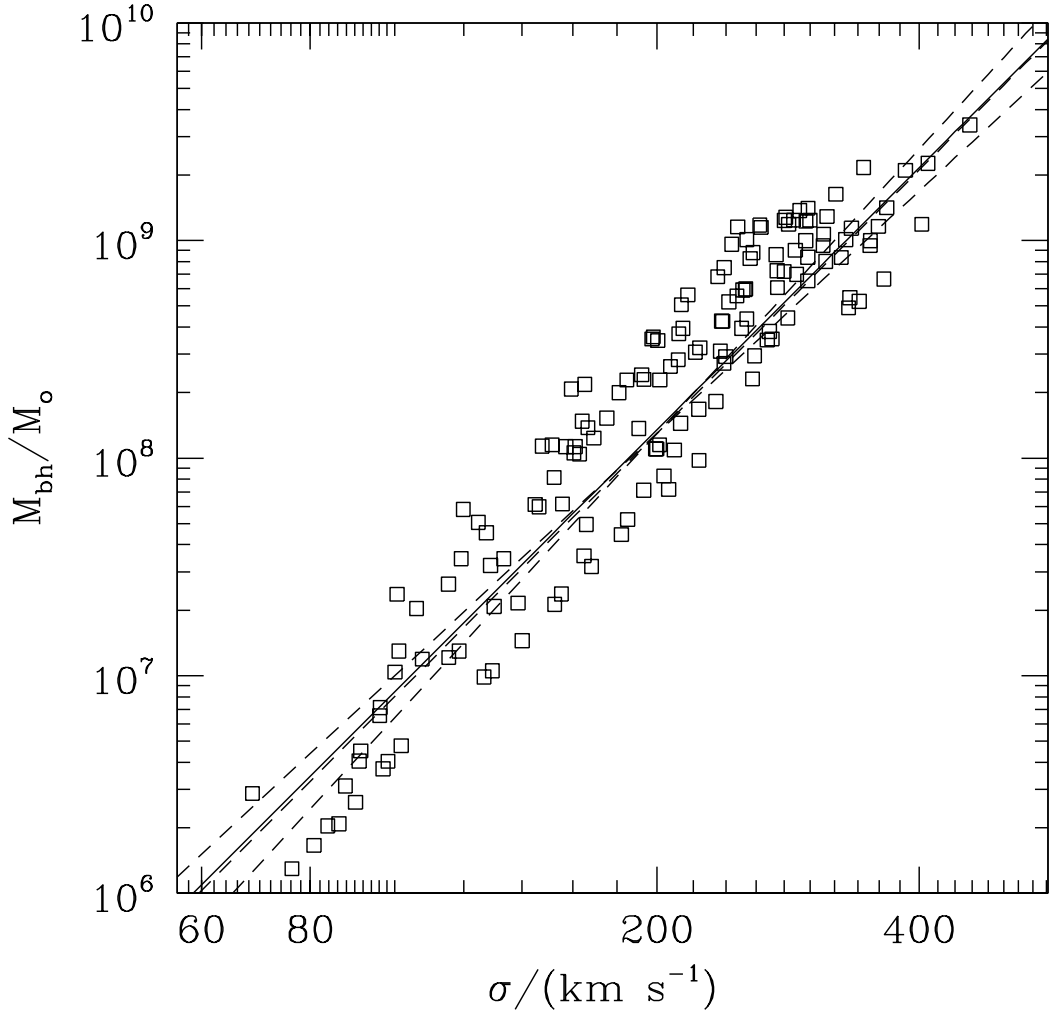}
\figcaption{An illustration of the correlation between black hole mass
$\mbh$ and velocity dispersion $\sigma$ being preserved under the
action of mergers. The result of each numerical experiment is shown
as an open square. The lines (for reference) are the same as those in
Figure 1. The numerical experiments begin with $N=2-6$ smaller units, 
which are merged according to the rules of \S 3.5. }  
\end{figure}

\end{document}